\documentclass[twocolumn,showpacs,preprintnumbers,amsmath,amssymb]{revtex4}
\usepackage{graphicx}% Include figure files
\usepackage{dcolumn}% Align table columns on decimal point
\usepackage{bm}% bold math
\usepackage{color}
\usepackage{float}

\def\nl{\noindent}

\begin{document}

\title{Bounds on $Z^\prime$ from 3-3-1 model at the LHC energies}

\author{Y. A. Coutinho\footnote{Now at CERN, Switzerland}} 
\email{yara.amaral.coutinho@cern.ch}
\author{V. Salustino Guimar\~aes} 
\affiliation{
Instituto de F\'isica - Universidade Federal do Rio de Janeiro\\
Avenida Athos da Silveira Ramos, 149
Rio de Janeiro - Rio de Janeiro, 21941-972, Brazil
}

\author{A. A. Nepomuceno}%
\email{andre.asevedo@cern.ch}
\affiliation{RFM - PURO - Universidade Federal Fluminense \\
Rua Recife, s/n, Rio das Ostras, Rio de Janeiro,  28890-000, Brazil
}%

\date{\today}

\begin{abstract}

The Large Hadron Collider will restart with higher energy and luminosity in 2015.
This achievement opens the possibility of discovering new phenomena hardly described by
the standard model, that is based on two neutral gauge bosons: the photon and the $Z$.
This perspective imposes a deep and systematic study of
models  that predicts the existence of  new neutral gauge bosons. One of these models is
based on the gauge group $SU(3)_C \times SU(3)_L \times U(1)_N$ called the 3-3-1 model for short.

In this paper we perform a study with  $Z^\prime$ predicted in two versions of the 3-3-1 model and
compare  the signature of this resonance in each model version.
By considering the present and future  LHC energy regimes, we obtain
 some distributions and
the total cross section for the process $p + p \longrightarrow \ell^{+} + \ell^{-} + X$.
Additionally, we derive lower bounds on
$Z^\prime$ mass from the latest LHC results. Finally we analyze the LHC
potential for discovering this neutral gauge boson at $14$ TeV center-of-mass energy.

\end{abstract}


\maketitle

%\linenumbers

\section{Introduction}

The search for new physics is one of the top priorities after a particle
consistent with the Higgs boson has been found at the Large Hadron Collider (LHC) \cite{ATL1, CMSQ}.
Although this discovery can elucidate the mass-generation mechanism, it
is still believed that the standard model (SM) is not the ultimate truth,
and that physics beyond it must exist at the TeV scale.
New phenomena are predicted in various alternative models and theoretical extensions from SM.
The existence of a new neutral current, called $Z^\prime$, is a common feature of most of these models. 

Among the models that have new physics content, the 3-3-1 model is the one that provides an
elegant answer to one of the modern intriguing questions, the problem of fermion families in nature. The model is built so that
anomalies cancel out when all families are summed over, so the family number must be a multiple of the color number.

The phenomenological consequences of the 3-3-1 model depend on its version.
The different versions of this model are a consequence of the characteristics of the $SU(3)$ matrices.
It is well known that two representations of the group generators can be simultaneously diagonalized.
This makes the charge operator dependent on the ratio between $\lambda_3$ to $\lambda_8$
matrix representations leading to different model versions. There is a version with an
extra neutral $Z^\prime$ and charged  $V^\pm$ and $U^{\pm \pm}$ gauge bosons carrying double
leptonic charge, called bileptons.
Moreover, in this version the $Z^\prime$ width can be large, and it is usually called the minimal version
of the model \cite{PIV, FRA}. There are two versions of the model where
there are no exotic charged quarks, one is called the right-handed neutrino version  \cite{RHN1, RHN2, RHN3, RHN4, RHN5} and the other
we call the \"Ozer version \cite{OZ1, OZ2}. For both, the $Z^\prime$ is a narrow resonance.
As we will discuss in the next section, the properties of the new neutral boson depend on the model
version, which is determined by the charge operator. Consequently, one needs to establish
phenomenological criteria to disentangle these versions by analyzing the production cross section
and some angular distributions that follow from each of them.

Several studies have been performed in order to derive bounds on the mass of new gauge bosons. These bounds 
come from either direct experimental searches or from phenomenological analysis using the available experimental data. 
In the universe of the 3-3-1 model, bounds on $M_{Z^\prime}$ were obtained from different analyses, such as 
the contribution from exotics to the oblique electroweak correction parameters ($S$, $T$ and $U$) \cite{LIU, SAS, OMR},
corrections to the $Z$-pole observables for arbitrary values of $\beta$ \cite{OCH, FRE, GUT},
the study of the energy region where perturbative treatment is still valid \cite{ALE},
$Z^\prime$ and exotic boson mass contributions to the muon decay parameters \cite{NGL, BEL},
the decay $\mu \rightarrow 3 \ e $  \cite{SHE},
and the contribution from neutral bosons  to the flavor changing neutral current (FCNC) 
\cite{VAN, DUM, TAE, LIUb, LIUc, JAI, SHERb, BEN, DUM2, CAB, COG}. 

In the original work from F. Pisano and V. Pleitez \cite{PIV}, a very restrictive bound 
on $Z^\prime$ mass was obtained ($M_{Z^\prime} > 40$ TeV) by considering the contribution from the $Z^\prime$ to the $K_S^0 - K_L^0$ mass difference.
More recently, a work from V. Pleitez {\it et al.} \cite{ANA}, based on additional contributions from a 
light scalar boson to FCNC, lowered the strong previous limit on the new neutral gauge boson for the minimal version 
of the 3-3-1 model. This new result allows the minimal 3-3-1 model predictions to be probed at LHC. 

Direct experimental searches performed by D\O\ \cite{D0} and CDF \cite{CDF1, CDF2} Collaborations derived bounds on 
$Z^\prime$ mass based on analyses with dielectron and dimuon final states at $\sqrt s = 1.96$ TeV.
They established lower bounds for different models and had excluded a $Z^\prime$ with mass in the range from $963$ to $1030$ GeV.

Recently, the ATLAS and CMS Collaborations presented results on narrow resonances with
dilepton final states
($e^+ e^-$ and $\mu^+ \mu^-$)  \cite{ATL2, ATL3, CMS1, CMS2}
and excluded a  sequential standard model  $Z^\prime$
with mass smaller than $2.49$ TeV (ATLAS) and $2.59$ TeV (CMS).
Although their data have been interpreted in terms of different scenarios for physics beyond
the SM, no limits on the $Z^\prime$ from the 3-3-1 model was derived from the latest LHC results.
The purpose of this article is to derive these unknown limits.

In this paper we consider the production and decay of the 3-3-1 $Z^\prime$ in
the process $p + p \longrightarrow \ell^{+} + \ell^{-} + X$ ($\ell = e, \, \mu$) for
different LHC energy regimes, when $Z^\prime$ is a narrow resonance as predicted in two
versions of the model, namely, the right-handed neutrino
model (RHN) \cite{RHN1, RHN2, RHN3, RHN4, RHN5} and the \"Ozer model \cite{OZ1, OZ2}.

Studies using CDF results have excluded a $Z^\prime$ from the RHN model with mass below $920$ GeV \cite{MART}. 
For the \"Ozer model, no limit on $Z^\prime$ mass has been derived so far.
A previous study on $Z^\prime$ at the ILC energies was made by one of the 
authors, where it  was possible to disentangle versions from the 3-3-1 model, 
considering the process $e^+ + e^- \longrightarrow \mu^+ + \mu^-$ and 
establishing from hadronic final states lower bounds on $M_{Z^\prime}$ 
with $95\%$ C. L. \cite{ELM}. The possibility to see signals from these 
models will considerably 
increased at the LHC running at $14$ TeV, a scenario that we also explore in this work. 

This paper is organized as follows: in Sec. II we describe the right-handed neutrinos and \"Ozer versions, 
highlighting the differences between them. In Sec. III we present the $Z^\prime$ width and the total cross section for the 
process investigated and for different $Z^\prime$ masses. In Secs. IV and V, we derive lower bounds on the $Z^\prime$ mass at $\sqrt s = 8$ 
and $\sqrt s = 14$ TeV and explore the LHC potential to find this new state at $14$ TeV. The conclusions are presented in 
Sec. VI. 

\section{Two versions of the 3-3-1 model}
The 3-3-1 model has many attractive features: among them, it is free from anomalies considering the number of fermion families
equal to the quantum number of color. The beginning is the electric charge operator that defines the version of the model,  
\begin{equation}
Q = T_3 - \beta \ T_8 + X I
\label {beta} 
\end{equation}
\noindent where the two generators $T_3$ and $ T_8$  satisfy the $SU(3)$ algebra, 
$I$ is the unit matrix, and finally, $X$ is the $U(1)$ charge. 

Depending upon the $\beta$ value, the charge operator determines the arrangement of the fields for he minimal version 
$\beta =  \sqrt 3$; $\beta = 1/ \sqrt 3$ leads to a model with right-handed neutrinos (RHN) and quarks with ordinary charges. Also another choice,  
$\beta = -1/\sqrt 3$ leads to a model without exotic charges.

We are interested in the two following versions: the right-handed neutrino version with  $\beta = 1/ \sqrt 3$ called here version I 
\cite{RHN1, RHN2, RHN3, RHN4},  and the version with  $\beta = -1/\sqrt 3$ \cite{OZ1, OZ2}, called version II.

Both versions present, besides the ordinary gauge bosons ($\gamma, \, Z, \, W^\pm $), neutral extra gauge bosons $Z^\prime$ and single 
charged bileptons $V^\pm$ and neutral one $X^0$, which carry a double lepton number. The heavy exotic quarks carry ordinary charges, $2/3$ for 
u-type and $-1/3$ for d-type. 

Each lepton family is arranged in triplets; the first two elements are the charged and the neutral lepton and the third element is a conjugate 
of the charged lepton or neutral lepton, depending on the $\beta$ factor. In order to cancel anomalies, the quarks are arranged 
in triplets and antitriplets (one family must be different from the other two).

The Higgs structure to give mass to all particles is composed of
three triplets ($\chi$, $\rho$, $\eta$), whose neutral fields develop nonzero vacuum expectation values, 
respectively, $v_{\chi}$,  $v_{\rho}$, and  $v_{\eta}$. To reproduce the SM phenomenology, 
a large scale is associated to the vacuum expectation value $v_\chi$, which gives mass to the exotic quarks and extra gauge bosons. 
Thus we have the conditions $v_\chi \gg v_\rho, v_\eta$, with $v_\rho^2 + v_\eta^2 = v_W^2= \left( 246 \right)^2$ 
GeV$^2$.  

The general Lagrangian for the neutral current involving only the $Z^{\prime}$ contribution is
\begin{eqnarray}
&&{\cal L}^{NC} =-\frac{g}{2 \cos\theta_W}\sum_{f} \Bigl[\bar
f\,  \gamma^\mu\ (g^\prime_V + g^\prime_A \gamma^5)f \, { Z_\mu^\prime}\Bigr],
\end{eqnarray}

\noindent
where $f$ are leptons and quarks, the couplings $g^\prime_V$ and $g^\prime_A$ 
are shown in Tables \ref{tab2} and \ref{tab1} for RHN and \"Ozer versions, $g$ is the $SU_L(3)$  
coupling, and $\theta_W$ is the Weinberg angle. 

\begin{table}
\begin{footnotesize}
\begin{center}
\begin{tabular}{|c|c|c|}
\hline
\multicolumn{3}{|c|}{RHN - version I} \\
\hline
&  $g^\prime_V$ & $g^\prime_A$   \\ 

\hline
$Z^{\prime} \bar \ell \ell $ &
$\displaystyle{\frac{-1+4\sin^2\theta_W}{2\sqrt{3-4\sin^2\theta_W}}}$ &
$\displaystyle{\frac{1}{2\sqrt{3-4\sin^2\theta_W}}}$ \\
\hline
$Z^{\prime} \bar u u$ & 
$\displaystyle{\frac{3-8\sin^2\theta_W}{{6\sqrt{3-4\sin^2\theta_W}}}}$  & 
$\displaystyle{-\frac{1}{2\sqrt{3-4\sin^2\theta_W}}}$  \\
\hline
$Z^{\prime} \bar d d$ &  
$\displaystyle{\frac{3- 2\sin^2\theta_W}{6\sqrt{3-4\sin^2\theta_W}}}$   &  
$\displaystyle{-\frac{{3-6\sin^2\theta_W}}{6\sqrt{3-4\sin^2\theta_W}}}$  \\
\hline
\end{tabular}
\end{center}
\end{footnotesize}
\caption{The vector and axial couplings of $Z^{\prime}$ with leptons ($e$, $\mu$, and $\tau$) and quarks ($u$ and $d$) 
in the RHN (version I). $\theta_W$ is the Weinberg angle.}
\label{tab2}
\end{table}

\begin{table}
\begin{footnotesize}
\begin{center}
\begin{tabular}{|c|c|c|}
\hline
\multicolumn{3}{|c|}{\"Ozer - version II} \\
\hline
&  $g^\prime_V$ & $g^\prime_A$   \\
\hline
$Z^{\prime} \bar \ell \ell $ &
$\displaystyle{-\frac{1+2\sin^2\theta_W}{2\sqrt{3-4\sin^2\theta_W}}}$ &
$\displaystyle{\frac{1-2\sin^2\theta_W}{2\sqrt{3-4\sin^2\theta_W}}}$ \\ %&
\hline
$Z^{\prime} \bar u u$ &
$\displaystyle{-\frac{3+2\sin^2\theta_W}{6\sqrt{3-4\sin^2\theta_W}}}$ &
$\displaystyle{\frac{1-2\sin^2\theta_W}{2\sqrt{3-4\sin^2\theta_W}}}$ \\ %& 
\hline
$Z^{\prime} \bar d d$ &
$\displaystyle{\frac{-3+4\sin^2\theta_W}{6\sqrt{3-4\sin^2\theta_W}}}$   &
$\displaystyle{\frac{1}{2\sqrt{3-4\sin^2\theta_W}}}$  \\% & 
\hline
\end{tabular}
\end{center}
\end{footnotesize}
\caption{The vector and axial couplings of $Z^{\prime}$ with leptons ($e$, $\mu$, and $\tau$) and quarks ($u$ and $d$) in the
\"Ozer (version II). $\theta_W$ is the Weinberg angle.}
\label{tab1}
\end{table}

\section{Numerical Implementation}

The two versions of 3-3-1 models discussed above were implemented in the COMPHEP package \cite{comp}, which was used for cross-section calculation 
and event generation. The parton distribution functions (PDF) CTEQ6L were used and the QCD factorization scale was set as the dilepton invariant mass of the event. 
Concerning the particle parameters, we considered heavy quarks, 
heavy leptons, and bileptons masses to be $1$ TeV, and we took the $Z^\prime$ mass in 
the range from $500$ to $4000$ GeV. 

\begin{figure}
\rotatebox{-360}{\scalebox{0.35}{\includegraphics{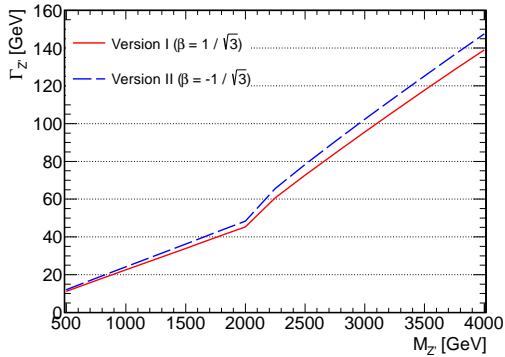}}}
\caption{$Z^\prime$ width as a function of $M_{Z^\prime}$ for versions I 
and II of the 3-3-1 model.}  
\label{fig1}
\end{figure}

In Fig. \ref{fig1} we present the total $Z^\prime$ width as a function of its mass for the two versions studied here. As we can see, the resonance 
is narrow in both versions, varying from $2\%$ to $4\%$ of $M_{Z^\prime}$ in the mass range considered. 
At $M_{Z^\prime} = 2$ TeV the slope of the curve increases because, from this point, the decay of $Z^\prime$ into exotic quarks becomes kinematically 
allowed. In both versions, the new neutral gauge boson can also decay into exotic 
fermions with branching ratios of order of $2\%$.

\begin{figure}[h!]
\begin{center}
\rotatebox{-360}{\scalebox{0.35}{\includegraphics{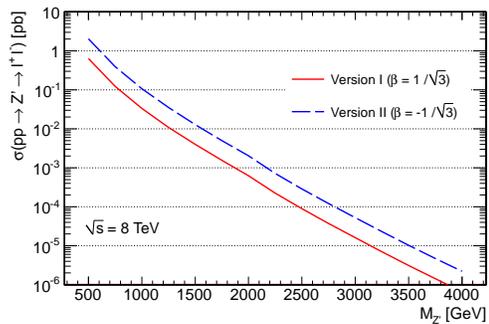}}}
\caption{Total cross section as a function of $M_{Z^\prime}$ 
for the process $p + p \longrightarrow Z^\prime \longrightarrow \ell^+ + \ell^-$
in versions I and II of the 3-3-1 model at $\sqrt s =8$ TeV. A cut
on the dilepton invariant mass of $M_{Z^\prime}$/2 was applied for this calculation.}  
\label{fig2}
\end{center}
\end{figure}

\begin{figure}[h!]
\begin{center}
\rotatebox{-360}{\scalebox{0.35}{\includegraphics{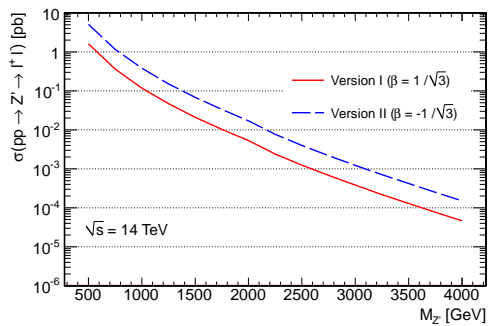}}}
\caption{Total cross section as a function of $M_{Z^\prime}$ for the 
process $p + p \longrightarrow Z^\prime \longrightarrow \ell^+ + \ell^-$
in versions I and II of the 3-3-1 model at $\sqrt s = 14$ TeV with
the same cut as Fig. \ref{fig2}}  
\label{fig3}
\end{center}
\end{figure}

Figure \ref{fig2} shows the total cross section calculated at tree level for the process 
$p + p \longrightarrow \ell^{+} + \ell^{-} + X$ at $\sqrt{s} = 8$ TeV,
where $\ell$ is either an electron or a muon. Figure \ref{fig3} shows the same cross section calculated for $14$ TeV. 
Both versions foresee cross sections that can to be probed at the LHC. Version II is the most optimistic since 
the $Z^\prime$ coupling to leptons is stronger than in version I. Note that depending on $M_{Z^\prime}$, 
the cross sections increase by a factor of $10$ to $10^2$ at $14$ TeV in comparison with their value at $8$ TeV.

\section{Exclusion limits at $\sqrt s=8$ TeV}

The LHC experiments have performed many analyses searching for signals of new spin $1$ gauge bosons in different final states, 
but so far no deviation of SM has been found. These analyses are usually model dependent, where a set of benchmark model
predictions are compared to data.

In the absence of any signal, ATLAS and CMS Collaborations have extended the $E_6$ superstring-inspired $Z^\prime$ exclusion mass 
to above $2$ TeV with $6$ fb$^{-1}$ and $4$ fb$^{-1}$ of collision data, respectively, at $\sqrt{s} = 8$ TeV \cite{ATL3, CMS2}. 
In particular, the CMS Collaboration has combined the results from $7$ and $8$ TeV to set $95\%$ C. L. limits on the 
ratio $R_{\sigma}$ of the cross section times branching fraction for $Z^\prime$ to that of the SM,

\begin{equation}
R_{\sigma}= \frac{\sigma(p + p \longrightarrow Z^\prime \longrightarrow \ell^+ + \ell^-)}{\sigma(p + p \longrightarrow Z \longrightarrow 
\ell^+ + \ell^-)}.
\end{equation}

We use the  CMS results to set lower limits on the $Z^\prime$ mass from 3-3-1 models. Following what was done by CMS, the $Z^\prime$ 
cross-sections for both versions are calculated in a range of $40\%$ about the $Z^\prime$ pole mass and the $Z$ cross-section is calculated 
in the interval $60$ GeV $<  m_{\ell \ell} < 120$ GeV. The ratio $R_{\sigma}$ is evaluated for $Z^\prime$ masses in the range between $500$ and $3000$ GeV. 

\begin{figure}[h!]
\rotatebox{-360}{\scalebox{0.35}{\includegraphics{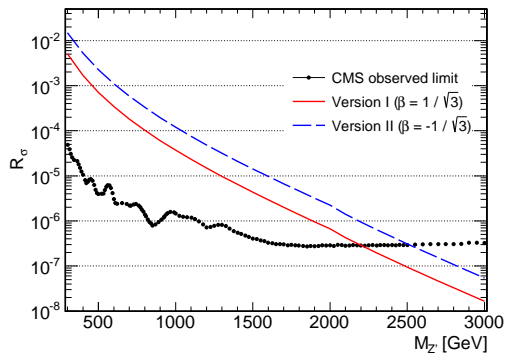}}}
\medskip
\caption{ $R_{\sigma}$ curves for both versions. The CMS observed limits on $R_{\sigma}$ (black dots) 
are for the combination of electron and muon channels at $7$ and $8$ TeV.}  
\label{fig4}
\end{figure}

Figure \ref{fig4} shows the CMS observed limits and the theoretical ratio $R_{\sigma}$ curve for both versions. The $Z^\prime$ lower mass limit is 
obtained from the point where the theoretical ratio curve crosses 
the observed limit. From the plot, we can conclude that the current data exclude with $95\%$  C. L. the version I new neutral 
gauge boson with mass below $2200$ GeV and the version II new resonance lighter than $2519$ GeV. This result does not change significantly 
if the value of exotic quark mass is changed. 

\section{Discovery potential and limits at $\sqrt{s}$ = 14 TeV }

After a shutdown in 2013 that is expected to take two years, the LHC will restart its operation at the 
design center-of-mass energy of $14$ TeV. Here we assume this scenario to investigate the LHC potential to find a 
$Z^\prime$ from the 3-3-1 model and determine the lower bounds on the $Z^\prime$ mass that can be set with this energy regime.

In order to determine the minimal integrated luminosity needed to claim a $Z^\prime$ discovery or to exclude it, the number 
of background and signal events expected in the processes $ p + p \longrightarrow e^{+} + e^{-} + X$ and 
$p + p \longrightarrow \mu^{+} + \mu^{-} + X$ are calculated. To make our results more realistic, we consider an overall efficiency of $66\%$ 
for the electron channel and $43\%$ for the muon channel, as determined by the ATLAS experiment \cite{ATL2}. These efficiencies 
take into account the geometrical acceptance of the detector ($\arrowvert \eta \arrowvert  < 2.5$), cuts on lepton transverse momentum
and lepton reconstruction and identification efficiencies. 

The dominant and irreducible background taken into account in this paper is the Drell-Yan (DY) process.
Although the $Z^\prime$ interfere with the $Z/ \gamma*$ process, the interference is minimal, and therefore we treat 
signal and background  as independent. Other backgrounds include QCD jets and ttbar events, but at high masses these 
backgrounds can be heavily suppressed by isolation cuts and are not considered here.  

\begin{figure}[h!]
\rotatebox{-360}{\scalebox{0.35}{\includegraphics{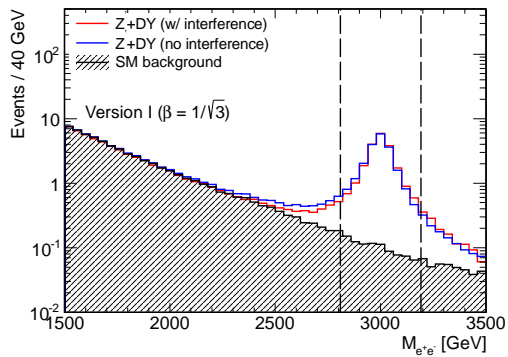}}}
\caption{Invariant mass distribution in the electron channel for $M_{Z^\prime} = 3000$ GeV for version I. The vertical lines 
represent the mass window used for selecting signal events.}  
\label{fig5}
\end{figure}

\begin{figure}
\rotatebox{-360}{\scalebox{0.35}{\includegraphics{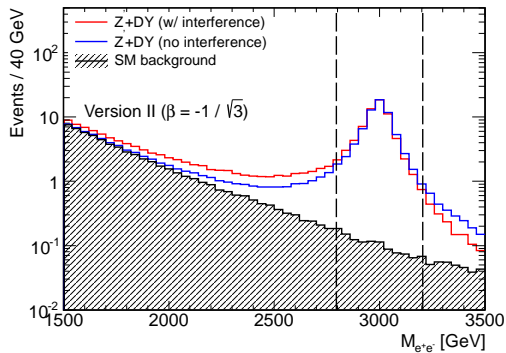}}}
\caption{Invariant mass distribution in the electron channel for $M_{Z^\prime} = 3000$ GeV for version II. The vertical lines 
represent the mass window used for selecting signal events.}  
\label{fig6}
\end{figure}

Figures \ref{fig5} and \ref{fig6} show the invariant mass distributions for the DY and for a signal mass hypothesis of 3000 GeV 
for  versions I and II, considering $100$  fb$^{-1}$ of data and the efficiencies mentioned above. Only the distribution for the 
electron channel is shown, but at the generator level the muon  channel distributions looks the same. To determine the 
significance of a signal such as those shown in the plots, we estimate the number of signal and background events by calculating 
the cross sections within a window $[M_{Z^\prime} -  2 \Gamma_{Z^\prime},  M_{Z^\prime} +  2 \Gamma_{Z^\prime}]$ for both channels. 
This selection, represented by the two vertical lines in Figs. \ref{fig5} and \ref{fig6}, suppresses considerably the
background while maintaining high signal efficiency. We can also see in these plots the effect of $Z^\prime$-DY interference 
on the invariant mass, which is small in both models, and under the selected mass window around the $Z^\prime$ mass, 
it is highly suppressed. 

The potential of the search to find a $ Z^\prime$ of a given mass is determined by the integrated luminosity needed 
to observe a signal with statistical significance of $5 \sigma$. The significance is obtained via the estimator \cite{cowan},

\begin{equation}
S = \sqrt{2((N_s+ N_b) \ln(1 + \frac{N_s}{N_b}) - N_s)}
\end{equation}

\noindent
\newline
where $N_s$ and $N_b$ are, respectively, the number of signal and backgrounds events expected in the mass window mentioned above. 

\begin{figure}[h!]
\rotatebox{-360}{\scalebox{0.35}{\includegraphics{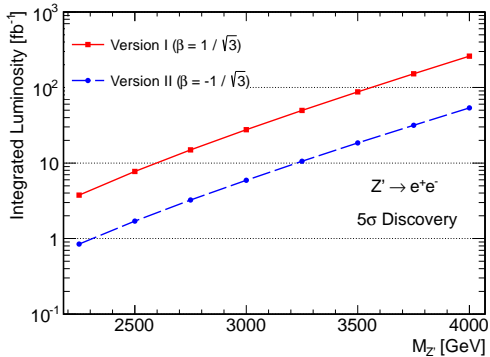}}}
\caption{Discovery potential  for versions I and II as a function of $M_{Z^\prime}$ at $14$ TeV in electron channel.}  
\label{fig7}
\end{figure}

\begin{figure}
\rotatebox{-360}{\scalebox{0.35}{\includegraphics{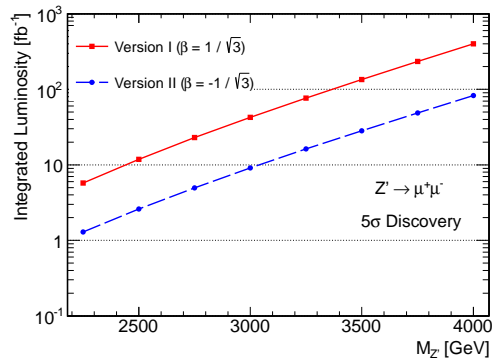}}}
\caption{Discovery potential for versions I and II as a function of $M_{Z^\prime}$ at $14$ TeV in muon channel.}  
\label{fig8}
\end{figure}

Figures \ref{fig7} and \ref{fig8} show the amount of integrated luminosity required to have a 5$\sigma$ $Z^\prime$ discovery in the 
electron and muon 
channels for both versions. As we can see, a 3-3-1 $Z^\prime$ with mass just above the exclusion limit ($2519$ GeV) 
can be reached with an amount of data of order of $1$ fb$^{-1}$ to $10$ fb$^{-1}$, depending on the channel and model. 
This scenario can be achieved in the first year of LHC operation at $14$ TeV. For $M_{Z^\prime} \sim 4$ TeV in version II, 
the amount of data required to discover this new heavy state would be less than $100$ fb$^{-1}$, while for version I, at least $250$ 
fb$^{-1}$ of data would be needed to observe a boson with that mass. 

If no resonance is found in the data, the current $Z^\prime$ limits can be considerably extended in the next years. 
Assuming the presence of only background in the data, we can calculate the expected limits on various $Z^\prime$ mass hypotheses 
considering different integrated luminosities. This is done by performing a single-bin likelihood analysis, using the estimated number 
of signal and background events and the algorithm described in \cite{junk}. It adopts a frequentist approach to compute the 
confidence level for exclusion of small signals by combining different searches. 
The electron and muon channels are combined to set $95\%$ C. L. exclusion on $\sigma \times Br (Z^\prime \longrightarrow \ell^+ + \ell^-)$, 
and these limits are translated to limits on $M_{Z^\prime}$. 

\begin{figure}[H]
\rotatebox{-360}{\scalebox{0.35}{\includegraphics{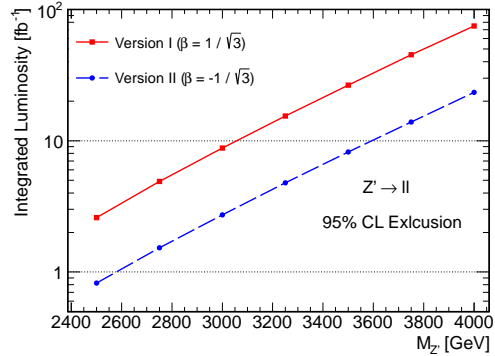}}}
\caption{Minimal integrated luminosity needed to exclude a $Z^\prime$ from version I and version II 
as a function of $M_{Z^\prime}$ at $14$ TeV.}  
\label{fig9}
\end{figure}

Figure \ref{fig9} shows the minimal integrated luminosity needed to exclude the new gauge boson as a function of $M_{Z^\prime}$. 
With $\sim 23$ fb$^{-1}$ of data, the version II $Z^\prime$ can be excluded up to masses of $4000$ GeV, but for version I, 
it would  need at least $3$ times more luminosity to exclude a $Z^\prime$ with mass of $4000$ GeV. 
Note that for $M_{Z^\prime} \sim 3000$ GeV, less than $10$ fb$^{-1}$ of data is enough for exclusion. This is important to point out 
because, although we have not considered in this work the 3-3-1 version that has theoretical upper bounds on $Z^\prime$, 
our results suggest that such a version can be completely excluded in the very early stages of LHC running at $14$ TeV,
since these upper bounds are usually below $3500$ GeV. 

\section{Conclusions}
New resonances are expected to manifest at LHC in the next years, and among them, the neutral heavy gauge boson $Z^\prime$ has 
a special role since it appears in different beyond-SM scenarios. In this paper we have presented a study involving the 3-3-1 model 
predictions, considering the process $p + p \longrightarrow \ell^{+} + \ell^{-} + X$. Lower limits on $Z^\prime$ 
mass from two versions of the 3-3-1 model were derived using the latest CMS published results. For the RHN model, a $Z^\prime$ with mass 
below $2200$ GeV is excluded. This limit is a considerable improvement of the bounds obtained with CDF results. 
On the other hand, we derived a first limit for the \"Ozer version: a $Z^\prime$ lighter than $2519$ GeV is excluded. 

Considering the LHC running at the design center-of-mass energy of $14$ TeV, we have shown that
a new resonance with mass of $4000$ GeV can be reached at LHC with integrated luminosities of order of $100$ fb$^{-1}$. On the other hand, 
if no signal is found, the LHC can already exclude $M_{Z^\prime} = 4000$ GeV in the first year of operation at the high-energy regime. 
This is the first investigation of this kind performed for the 3-3-1 models considering the LHC upgraded energy. 
As the 3-3-1 model predicts a number of new particles, the observation of a $Z^\prime$ in combination 
with other exotic searches like bileptons and leptoquarks would provide a powerful way of discriminating between 3-3-1 versions and other 
BSM scenarios with new neutral heavy states.

\vskip 1.5cm
\nl {\bf Acknowledgments}

The authors would like to thank Prof. Jos\'e de S\'a Borges from UERJ (RJ - Brazil) for helpful suggestions.

Y.~A.~Coutinho thanks CNPq and FAPERJ,  V.~S.~Guimar\~aes thanks CAPES and A.~A.~Nepomuceno thanks FAPERJ for financial support.

\end{document}